\documentclass[
    aps,
    pra,
    twocolumn,
	colorlinks=true,
	linkcolor={teal},
	citecolor={teal},
    urlcolor={teal}, 
    nofootinbib,
    groupedaddress
]{revtex4-2}

\usepackage{orcidlink}
\usepackage{amsmath,amsfonts}
\usepackage{amssymb}
\usepackage{cancel}
\usepackage{dsfont}
\usepackage{pifont}

\usepackage{xpatch}
\usepackage{enumitem}
\usepackage{algorithm}
\usepackage{varwidth}
\usepackage[indLines=false]{algpseudocodex}
\makeatletter
\xpatchcmd{\algorithmic}{\itemsep\z@}{\itemsep=0.5ex}{}{}
\makeatother

\usepackage{array}
\usepackage{physics}
\usepackage{textcomp}
\usepackage{stfloats}
\usepackage{url}
\usepackage{verbatim}
\usepackage{graphicx}
\usepackage{braket}
\usepackage[capitalize,noabbrev]{cleveref}
\usepackage{todonotes}
\usepackage{quantikz}
\usepackage[dvipsnames]{xcolor}
\usepackage{csquotes}
\usepackage[normalem]{ulem}
\usepackage[acronym]{glossaries}
\glsdisablehyper
\usepackage{tikz}
\usetikzlibrary{
    quantikz2,
    graphs,
    quotes,
}

\usepackage{pgfplotstable}
\usepackage{pgfplots}
\pgfplotsset{compat=1.3}
\usepgfplotslibrary{statistics}
\usepgfplotslibrary{fillbetween}
\usepgfplotslibrary{groupplots}

\newcommand{\matr}[1]{{\boldsymbol{#1}}}
\renewcommand{\vec}[1]{{\boldsymbol{#1}}}

\newcommand{\Dist}{\mathcal{D}}

\newcommand{\Bd}{\mathfrak{B}}
\newtheorem{theorem}{Theorem}

\newacronym{hhl}{HHL}{ Harrow–Hassidim–Lloyd}
\newacronym{ccbqp}{CC-BQP}{cardinality constraint binary quadratic optimization problem}
\newacronym{qubo}{QUBO}{quadratic unconstrained binary optimization}
\newacronym{ae}{AE}{amplitude estimation}
\newacronym{aa}{AA}{amplitude amplification}
\newacronym{qram}{QRAM}{quantum random access memory}
\newacronym{qpe}{QPE}{quantum phase estimation}
\newacronym{ouralgo}{QSSA}{quantum subgraph similarity algorithm}

\begin{document}
\title{Quantum search algorithm for similar subgraph identification under fixed edge removal
}
\author{Ruben Kara \orcidlink{0000-0001-9258-4397}}
\email{r.kara@fz-juelich.de}
\affiliation{Institute of Climate and Energy Systems (ICE-1), Forschungszentrum Jülich, Jülich, Germany}
\author{Sven Danz \orcidlink{0009-0008-5083-3920}}
\affiliation{Institute for Quantum Information, RWTH Aachen University, Aachen, Germany}
\author{Tobias Stollenwerk \orcidlink{0000-0001-5445-8082}}
\affiliation{Institute for Quantum Computing Analytics (PGI-12), Forschungszentrum Jülich, Germany}
\author{Andrea Benigni \orcidlink{0000-0002-2475-7003}}
\affiliation{Institute of Climate and Energy Systems (ICE-1), Forschungszentrum Jülich, Jülich, Germany}
\affiliation{Computational Methods for Energy Systems Engineering, RWTH Aachen University, Aachen, Germany}

\begin{abstract}
\noindent
We introduce a novel quantum algorithm for similar subgraph identification in form of an NP-hard cardinality-constrained binary quadratic optimization problem. 
Given a weighted reference graph with Laplacian $\matr B$, our algorithm determines the subgraph featuring Laplacian $\matr B'$ on the same vertex set, but $x$ out of $N$ inactive edges, minimizing the Frobenius distance $||\matr B - \matr B'||_\mathrm{F}^2$. 
We represent the $\binom{N}{x}$ graph topologies by an equal-weight superposition in form of a Dicke state, enabling controlled transformations applied to the quantum state associated with the vectorized Laplacian of the reference graph. 
Combined with amplitude estimation and a minimum finding approach, our algorithm provides a polynomial speed up $\mathcal{O}(\sqrt{N^{x}/x!}N\log\log N)$ compared to $\mathcal{O}(N^{x+1}/x!)$ of classical brute-force search algorithms. 
We demonstrate the application of our method on standard test cases, which represent electric power grids, by reconstructing $||\matr B - \matr B'||_\mathrm{F}^2$ from measurements and show how our approach can be additionally used to calculate energy functional like quadratic forms of the Laplacians with respect to a given vector.    
\end{abstract}

\maketitle

\section{Introduction}
\label{sec:intro}
\noindent
Many quantum algorithms promise quadratic or even exponential advantages in terms of time or space complexity compared to the best known classical approaches. 
These algorithms typically rely on idealized access models and cost assumptions for primitives such as Hamiltonian simulation whose overhead can dominate the overall complexity at the end.
Therefore, in order to retain a potential advantage in an end-to-end solution for a particular application, a careful consideration of classical data input and output, as well as implementation details are imperative.
From this perspective, our work is motivated by the $N$-$x$ contingency analysis problem, typically arising in the context of power grid operation, but may be applied to related network problems as well. 
In particular, the $N$-$x$ contingency analysis addresses the stability, reliability and resilience of a network in the case of $x$ failing components out of $N$ total components providing the large number of
\begin{equation}
    S
    =
    \binom{N}{x}
    =
    \mathcal{O}\left(
        \frac{N^x}{x!}
    \right)
\end{equation}
different scenarios to analyze. 
This real-world problem is both relevant and computational demanding: In modern power grids, extreme weather events increase the challenges for the reliability of critical infrastructure~\cite{Montoya-Rincon,entso}, whereby the exhaustive simulation of realistic contingency sets remains challenging despite recent advances in exascale computing (\href{https://www.llnl.gov/article/50066/llnl-team-reaches-milestone-power-grid-optimization-worlds-first-exascale-supercomputer}{llnl-team-reaches-milestone-power-grid-optimization-worlds-first-exascale-supercomputer}).
The contingency analysis problem can be seen as a two-fold problem. 
First, given a particular grid configuration the power flow across the network needs to be simulated accurately.
In this context, the \gls{hhl} algorithm \cite{harrow2009quantum} has been considered for power flow problems~\cite{feng2021,Liu:2022rxb,feng2023}.
However, as \gls{hhl} comes with challenges which need to be considered in order to retain any quantum advantage~\cite{Aaronson}, it remains unclear if such approaches could lead to a scalable end-to-end advantage.

Therefore, we focus on the second part of the contingency analysis problem:
Instead of trying to accelerate the power flow simulation of each individual contingency scenario using quantum routines such as \gls{hhl}, we focus on the combinatorial explosion in the number of scenarios by searching over graph topologies. In particular, we aim to use a quantum algorithm to determine the subgraph out of $S$-many featuring $x$ edges removed, that is the most similar to the reference graph, 
where all edges are intact.
More precisely, given an undirected graph $G=(V, E)$ with $M$ nodes and $N$ edges, 
as well as positive edge weights
$\{b_e \in \mathbb{R}^+ \mid e \in E \}$,
we strive to find subgraphs of $G$ sharing the same vertex set $V$, with $x$ out of $|E|=N$ edges removed, that are most similar to the original graph. 
For this purpose, a binary decision variable $\tilde{d}_i \in \{0, 1\}$ for $i \in E$ indicates if an edge remains.
Hence, each binary vector $\vec{\tilde{d}} \in \{ 0,1\}^N $ represents a subgraph with a number of edges removed. We refer to these also as \emph{grid topologies} or \emph{configurations}.
If $\Dist(\vec{\tilde{d}})$ denotes a measure for the similarity of that subgraph to the original graph, 
our cost function reads
\begin{equation}\label{eq:distance-definition}
    \min_{\vec{\tilde{d}}} \Dist (\vec{\tilde{d}})\qquad \text{s.t. } \abs{\vec{\tilde{d}}}^2 =\sum_{i=0}^{N-1} \tilde{d}_i=N-x.
\end{equation}

We denote the Laplacian  of the \emph{reference} configuration with all edges activated as $\matr B \in \mathbb{R}^{M \times M}$ whose elements can be written in the node space as
\begin{align}
    B_{rs} 
    = 
    \begin{cases}
        \sum_{l\in \mathcal{N}_r} b_{rl}  & \; \mbox{if } \, r = s, 
        \\
        - b_{rs} & \; \mbox{if $r$ and $s$ are adjacent},  
        \\
        0     & \; \mbox{otherwise}.
    \end{cases} 
    \label{eq:laplacian_definition}
\end{align}
where $\mathcal{N}_r =\{l \in V\mid (r,l)\in E\}$ denotes the neighborhood of node $r$
and $b_i=b_{rs}$, if edge $i=(r,s)$ connects node $r$ and node $s$. 
Given a configuration $\vec{\tilde{d}}$, the elements of the corresponding Laplacian read
\begin{align}
     B_{rs}^\vec{\tilde{d}}
     =  
     B_{rs} \tilde{d}_{rs},
\end{align}
with the binary $\tilde{d}_{rs}=\tilde{d}_{sr}\in\{\tilde{d}_i\mid \forall i \in E \}$ which indicates whether the edge $(r,s)$ - originally connecting node $r$ and node $s$ in the reference graph - is active or not in the subgraph associated with configuration $\vec{\tilde{d}}$.
Note that we can write the Laplacian in the edge space as
\begin{align}
    \matr B= \sum_{i=0}^{N-1} b_i \vec{v}_i^T \vec{v}_i
\end{align}
where $\vec{v}_i$ is the $i$-th column vector of the underlying incidence matrix (see \cref{sec:methods_vectorizing}). As a measure for the similarity $\Dist(\vec{\tilde{d}})$, we use the squared Frobenius distance between the reference Laplacian $\matr B$ and the subgraph Laplacian $\matr B^\vec{\tilde{d}}$. Thus, we have
\begin{multline}
    \min_{\vec{\tilde{d}}} \Dist(\vec{\tilde{d}})
    =
    \min_{\vec{\tilde{d}}} ||\matr B- \matr B^\vec{\tilde{d}}||_\mathrm{F}^2 
    \\
    =
    \min_{\vec{\tilde{d}}}  \left[ \sum_{i=0}^{N-1} \sum_{j=0}^{N-1} b_i b_j (1-\tilde{d}_i)(1-\tilde{d}_j)(\vec{v}_i^T \vec{v}_j)^2 \right]
    \\
    =
    \min_{\vec{\tilde{d}}} \left[ 
        \vec{\tilde{d}}^T \matr Q \vec{\tilde{d}} 
        - 2 \sum_{i=0}^{N-1} \sum_{j=0}^{N-1}Q_{ij} \tilde{d}_i  
    \right]
    \\
    =
    \min_{\vec{\tilde{d}}} \left[ (\vec{\tilde{d}} -\vec{1})^T \matr Q (\vec{\tilde{d}} -\vec{1}) \right]
    =
    \min_{\vec{d}}\; \vec{d}^T \matr Q \vec{d},
    \label{eq:frob_class} 
\end{multline}
with the condition
\begin{align}
     \sum_{i=0}^{N-1} d_i=x ,
\end{align}
which is equivalent to $ \sum_{i=0}^{N-1} \tilde{d}_i=N-x$ by defining  the binary vector $\vec{d}=\vec{1}-\vec{\tilde{d}}$. The elements of the real and symmetric matrix $\matr Q$ read
\begin{align}
    Q_{ij}=b_i b_j (\vec{v}_i^T \vec{v}_j)^2.
\end{align}

This is a \gls{ccbqp} which is NP-hard with applications for instance, in the densest k-subgraph problem \cite{10447398}. 
Such NP-hard \gls{ccbqp} problems appear in many real-world applications and
near-term quantum computation solution have been considered where the cardinality constraint $|\vec{d}|^2=x$ is incorporated in the cost function as quadratic penalty term using the \gls{qubo} formulation \cite{venturelli2019reverse,stollenwerkATM2019}, or by  
restricting the search space to constraint-preserving solutions throughout the algorithm~\cite{hen2016quantum,hadfield2019from,stollenwerk_toward_2020}.

In this work, we introduce a novel quantum algorithm, called \gls{ouralgo}, 
which solves the above problem~\eqref{eq:frob_class}, i.e., identifies the subgraphs with fixed reduced cardinality $x$ which is most similar to a reference graph.
By associating the Laplacian of the reference graph with a quantum state and applying unitary transformations controlled by the Dicke state (equal superposition of fixed cardinality states), the algorithm generates a superposition of quantum states whose amplitudes correspond to the Frobenius distance up to a global factor. 
Adapting Dürr and H\o yer's minimum finder~\cite{durr1996quantum} and \gls{ae}~\cite{brassard2000quantum}, we show how this algorithm determines the configuration $\vec{\tilde{d}}$ minimizing the Frobenius distance and reconstruct exemplary $\Dist(\vec{\tilde{d}})$ from measurements to demonstrate the validity of our approach. 
Additionally, we show how our algorithm can be used to calculate quadratic forms $\vec{a}^T \matr B^\vec{\tilde{d}} \vec{a}$ for $\vec{a}\in \mathbb{R}^{M}$.

The determination of the most similar graph using graph similarity measures such as the Frobenius distance is an essential task with broad applications: Based on the Laplacians, they are used to study patterns of information transmission over networks using diffusion/heat-kernel distances \cite{hammond}, in machine learning for kernel alignement studies \cite{kornblith2019similarity}  or for spectral clustering \cite{andreotti2021measuring}.
More generally, identifying the nearest graph under a Frobenius distance between Laplacians has been studied as a standalone problem \cite{sato2024nearest,gervens_et_al:LIPIcs.MFCS.2022.52}. In the context of power systems, determining the closest graph Laplacian using the Frobenius distance is a common task in topology and state estimation \cite{grotas} or event localization such as switching actions and ground faults \cite{ardakananin} using the admittance matrices in the alternating current model. 

The remainder of this work is structured as follows.
We begin by presenting our main result, the \gls{ouralgo} in \cref{sec:results} including a complexity analysis, numerical proof-of-concept and application to the calculation of quadratic forms.
After a discussion of our results in \cref{sec:discuss} we present the algorithm details in \cref{sec:methods}.

\section{Results}
\label{sec:results}
In this section, we present and discuss our main result which is the \gls{ouralgo} leading us to the following proposition

\begin{theorem}
    \label{the:complexity}
    Let $G$ be a graph with $N$ weighted edges from which we form all subgraphs by deactivating a fixed number $x$ of edges.
    The configurations of activated edges and with it the subgraphs are denoted $\vec{\tilde{d}}$.
    There exists a quantum algorithm with internal precision $\log \varepsilon^{-1}$, that finds the configuration $\vec{\tilde{d}}$ with minimum Frobenius distance to the original graph in at most 
    \begin{equation}
        t_\mathrm{min}
        =
        \mathcal{O}\left(
            \sqrt{\frac{N^x}{x!}}
            \frac{N\log\log N}{\varepsilon}
        \right)
        \label{eq:runtime}
    \end{equation}
    steps using at most
    \begin{equation}
        n_\mathrm{min}
        =
        \mathcal{O}\left(
            \mathrm{min}(N, x\lceil\log N\rceil)
            + \log \frac{1}{\varepsilon}
        \right)
        \label{eq:memory}
    \end{equation}
    qubits.
\end{theorem}

The proof of the latter is split into two parts.
First we describe the algorithm briefly in \cref{sec:algorithm} -- the details follow in \cref{sec:methods}.
The second part, is its complexity analysis that we give in \cref{sec:complexity}.
We conclude with a numerical proof-of-concept in \cref{sec:numerics}.
Finally, we allude to the application of our approach to the calculation of quadratic forms in \cref{sec:quad_from}.

\subsection{Algorithm}
\label{sec:algorithm}
The algorithm, that satisfies \cref{the:complexity} is our \gls{ouralgo}.
Our main idea, here, is the parallel computation of the Frobenius distance $\mathcal{D}(\vec{\tilde{d}})$, for all subgraphs of $G$, in a quantum superposition.
Once encoded in the base of a quantum register we can utilize a quantum search algorithm to compare all subgraphs and return the configuration $\vec{\tilde{d}}$ with minimum $\mathcal{D}(\vec{\tilde{d}})$.
The quadratic quantum search speed up is the key that allows to compare the vast number of subgraphs efficiently and leads to \cref{eq:runtime}.

As just indicated, our \gls{ouralgo} can be separated into the computational part and the search part.
For the computation of the Frobenius distance, we start with encoding the configurations $\vec{\tilde{d}}$ of all subgraphs in form of a Dicke state $\ket{D_x^N}$ (cf. \cref{algo:minimum-subgraph-similarity}).
We describe another encoding in \cref{sec:topo-enc}, that is beneficial for small $x$ but will focus in this manuscript on the first for a better readability. 
Next, we need to encode the weights $b_i$, in the amplitudes of another quantum superposition $\ket{\vec{b}}$ with the edge labels $i$ as base.
This allows us to flag all inactive edges states, based on the configuration $\vec{\tilde{d}}$ represented by computational basis state $\ket{d}$ with Hamming weight $x$ as part of the Dicke state $\ket{D^N_x}$, with multi controlled operations.
Finally, we encode the column vectors $\vec{v}_i$ of the incidence matrix by applying a block encoding of the latter to the edge label register returning the final state $\ket{\psi_\mathrm{f}}$.
With this, we prepare the superposition $\ket{\psi_\mathrm{f}}$ whose amplitudes and probabilities to measure encode the Frobenius distances \cref{eq:frob_class}. Estimating those probabilities from brute force sampling is a slow process, which is why the second part of our \gls{ouralgo} contains a quantum minimum finder~\cite{durr1996quantum}. In particular, we use \gls{ae} to map the amplitudes related to the Frobenius distances to a \emph{label} register. From here, we can start with a random guess $\vec{y}$ for the configuration and compute the corresponding $\mathcal{D}(\vec{y})$, either classically or following the previous steps.
After this, we compare the two values and mark every configuration in the quantum superposition with $\mathcal{D}(\vec{\tilde{d}}) <\mathcal{D}(\vec{y})$ to amplify them with a quantum search algorithm.
Finally, we measure with high probability a configuration with a Frobonius distance, that is smaller than $\mathcal{D}(\vec{y})$.
Hence, we replace $\vec{y}$ with the measured one and repeat everything until we reach a small enough value.
The whole process is gathered in \cref{algo:minimum-subgraph-similarity}.

\begin{algorithm}[H]
\begin{minipage}{0.9\linewidth}
\textbf{Input:}
\begin{itemize}[leftmargin=2em]
    \item
        Graph $G=(V, E)$ with positive edge weights
        $\mathbf{b} :=\{b_e \in \mathrm{R}^+ \mid e \in E \}$
    \item 
        Number of edge removals $x > 0$
\end{itemize}
\textbf{Output:}
\begin{itemize}[leftmargin=2em]
    \item
        Configuration 
        $\tilde{d}_i \in \{0, 1\}$ for $i \in E$ 
        which minimized the distance according to \cref{eq:distance-definition}
\end{itemize}
\begin{algorithmic}[1]
    \State 
    $\Vec{y} \gets$
    randomly chosen from $[0,S]$
    \While{$\dots$} 
        \State 
            $
            \ket{D_x^N} 
            \xleftarrow[]{\eqref{eq:dicke_short}}
            G, x
            $
            \Comment{Encode graph topology}
        \State 
            $
            \ket{\mathbf{b}}
            \xleftarrow[]{\eqref{eq:weight-encoding}}
            \mathbf{b}, 
            $
            \Comment{Encode weights}
        \State 
            $
            \ket{\psi_\mathrm{f}}
            \xleftarrow[]{\eqref{eq:psi2}}
            \ket{\mathbf{b}}
            \ket{D_x^N} 
            $
            \\
            \Comment{Deactivate edges and encode Laplacian}
        \State 
            $
            \ket{\psi_\mathrm{label}}
            \xleftarrow[]{\eqref{eq:labeling}}
            \ket{\psi_\mathrm{f}}
            $ \Comment{Distance labeling via \acrshort{ae}}
        \State 
            $\ket{y} \gets \Vec{y}$
        \State 
            $
            \ket{\psi_\mathrm{label}^y}
            \xleftarrow[]{\eqref{eq:weight-encoding}, \eqref{eq:psi2}, \eqref{eq:labeling}}
            \Vec{b},\ket{y}
            $ \Comment{repeat step 5-7}
        \State Mark states with $\mathcal{D}(\vec{\tilde{d}}) <\mathcal{D}(\vec{y})$.
        \State Apply quantum search algorithm
        \State $y_\mathrm{meas}$ $\gets$ Measure $\ket{\psi_\mathrm{label}}$
        \If{$\mathcal{D}(\vec{y}_\mathrm{meas}) <\mathcal{D}(\vec{y})$}
            \State $\vec{y} \gets \vec{y}_\mathrm{meas}$ \Comment{replace best value}
        \EndIf
    \EndWhile 
    \Return $\vec{y}$%
\end{algorithmic}
\end{minipage}

\caption{Algorithm for similar subgraph identification}
\label{algo:minimum-subgraph-similarity}
\end{algorithm}

\subsection{Complexity analysis}
\label{sec:complexity}
We analyze the computational complexity of \cref{algo:minimum-subgraph-similarity} discussed in the previous section, with respect to two parameters -- the total runtime and the quantum memory requirements.
This will be limited to $\mathcal{O}$ notation, which focuses on the scaling with critical parameters.
Those are the edge number $N$, the vertex number $M$, the number of inactive edges $x$, the accuracy of the internal precision $\log \varepsilon^{-1}$, and the number of ancilla qubits $a_V$ required for the block encoding.

\subsubsection{Runtime analysis}
We analyze the runtime from outside to inside, which means we start with the minimum finder introduced in Ref.~\cite{durr1996quantum}.
The dominant subroutines of that algorithm are the \gls{ae} for the state preparation, the greater than operation $U_>$ for the comparison, and  the \gls{aa} as search algorithm.

The minimum finder stops after a total of $\mathcal{O}(\sqrt{S})$ steps, where we treat one iteration of the search algorithm as one step.
For a more detailed analysis we assume that we require at most $S_1\leq \sqrt{S}$ iterations of the minimum finder and $S_2\leq \sqrt{S}$ iterations in the \gls{aa}.
Hence the runtime is
\begin{equation}
    t_\mathrm{min} 
    = 
    \mathcal{O}\left(
        S_1\left(
            t_\mathrm{AE}
            +t_<
            +t_\mathrm{AA}
        \right)
    \right).
\end{equation}
$U_>$ can be implemented with just two quantum adders that scale at most $\mathcal{O}(\log\varepsilon^{-1})$~\cite{draper2004, cuccaro2004,vanmeter2005, thomsen2008} in both memory and runtime.
This is much faster than the \gls{ae} which is why we neglect it in the following complexity analysis.
The \gls{aa} requires access to the state preparation of $\ket{\psi_\mathrm{label}}$ via \gls{ae}.
Hence, the runtime requirements of the \gls{aa} scales like 
\begin{equation}
    t_\mathrm{AA}
    =
    \mathcal{O}(S_2 t_\mathrm{AE}).
\end{equation}
The minimum finder is designed to stop after $\mathcal{O}(\sqrt{S})$ steps, which translates to $S_1S_2 = \mathcal{O}(\sqrt{S})$.
This simplifies the total runtime of the minimum finder and we have
\begin{equation}
    t_\mathrm{min} 
    = 
    \mathcal{O}\left(
        \sqrt{S}
            t_\mathrm{AE}
    \right).
\end{equation}

The \gls{ae} requires the generation of the Dicke states followed by a \gls{qpe} of $\mathrm{c}Q$ up to an accuracy of $\varepsilon^{-1}$.
Here, $\mathrm{c}Q$ is a combination of precisely chosen reflections whose eigenphases encode the Frobenius distance (cf. \cref{sec:amp-amp}).
The runtime of the \gls{ae} is therefore given by
\begin{equation}
    t_\mathrm{AE}
    =
    \mathcal{O}\left(
        t_\mathrm{DS}
        + \frac{1}{\varepsilon}t_{\mathrm{c}Q}
    \right),
\end{equation}
where $t_\mathrm{DS}=\mathcal{O}(N)$~\cite{bartschi2019deterministic} denotes the encoding time of the $S$ Dicke states, and $t_{\mathrm{c}Q}$ the runtime of one iteration of $\mathrm{c}Q$.
The dominant routines in $\mathrm{c}Q$ are multi-controlled $Z$ gates, and the controlled generator of $\ket{\psi_\mathrm{f}}$ denoted with $\mathrm{c}U_\mathrm{f}$.
The runtime of $n$-controlled gates require usually $\mathcal{O}(n)$ elementary operations if the internal Toffoli gates connect the controls in series~\cite{barenco1995elementary}.
If geometrically possible, a cascade like connection yields a $t^{(n)}_\mathrm{Toffoli}=\mathcal{O}(\log n)$ runtime~\cite{heDecompositionsNqubitToffoli2017}.
In our case we need multi-controlled Z gates with $2a_V+1 $ controles.
The implementation of $\mathrm{c}U_\mathrm{f}$ consists of multiple steps that we analyze in detail step-by-step.

The initial step is the amplitude encoding of the $N$ edge states $\{b_0,\dots,b_{N-1}\}$.
This requires a runtime of $t_\mathrm{enc}=\mathcal{O}(N)$ for $N$ independent values.
The second step is the topological-controlled deactivation of edges via $U_\mathrm{rme}$ introduced in \cref{sec:topo}.
The core of this routine requires $N$ multi-controlled NOT gates with $\lceil\log N\rceil+1$ controls each.
Their individual runtime is $t^{(\log N)}_\mathrm{Toffoli}=\mathcal{O}(\log \log N)$.
The last part is the block encoding of the sparse non-square incidence matrix with a runtime of $t_E = \mathcal{O}(N)$.
Here, we assume that the implementation of an oracle, that grants access to $\matr E$ with $2N$ non-zero entries (see \cref{sec:methods_vectorizing}), scales like $\mathcal{O}(N)$.
This yields a total runtime of 
\begin{multline}
    t_{\mathrm{c}Q}
    =
    \mathcal{O}\left(
        t^{(2a_V+1)}_\mathrm{Toffoli}
        + t_\mathrm{enc}
        + N t^{(\log N)}_\mathrm{Toffoli}
        + t_E
    \right)
    \\
    =
    \mathcal{O}\left(
        \log a_V
        + N \log \log N
    \right).
\end{multline}
Here, the contributions from the amplitude encoding and the block encoding are negligible compared to the edge deactivation.
Using Theorem 1 of Ref.~\cite{camps2022_fable}, there is an upper bound for the number of ancilla qubits $a_V$
\begin{multline}
    \log a_V
    \leq 
    a_V+m
    =
    2\max(n,m)+1
    \\
    \leq
    2\max(N,M)+1,
\end{multline}
so that their contribution is negligible for the runtime.
Furthermore, we work with sparse graphs so that $N\propto M$ and with it, we can simplify $\max(N,M)\approx N$ in the total runtime.
With this we have
\begin{equation}
    t_\mathrm{min}
    =
    \mathcal{O}\left(
        \sqrt{S}
        \frac{N\log\log N}{\varepsilon}
    \right).
\end{equation}
At last, as the total number of contingency scnarios scales as $S = \mathcal{O}(N^x/x!)$, we can derive \cref{eq:runtime}
\begin{equation*}
    t_\mathrm{min}
    =
    \mathcal{O}\left(
        \frac{N^{x/2+1}}{\sqrt{x!}}\frac{\log\log N}{\varepsilon}
    \right).
\end{equation*}

\subsubsection{Memory requirements}

The memory requirements of our approach consists of $N$ qubits needed for the Dicke states $\ket{d_x^N}$, or $x\lceil\log_2 N\rceil$ if we use the secondary method introduced in \cref{sec:topo-enc}.
Additionally, it requires $2n$ with $n=\lceil\log_2 N\rceil$  qubits for the edge states $\ket{i}$, $2a_E$ ancilla qubits for the block encoding of the incidence matrix, one qubit for the edge deactivation, and $\lceil \log_2 \varepsilon^{-1} \rceil$ qubits for the secondary register used to store the phase in the \gls{qpe}.
We know the total qubit number necessary for the block encoding~\cite{camps2022_fable}, which is
\begin{equation}
    n+a_E
    =
    m + a_V
    =
    2\left\lceil
        \log_2 \max(N,M)
    \right\rceil
    +1.
\end{equation}
Additionally, we require ancilla qubits for intermediate results in the block encoding, and the Toffoli gates in $U_\mathrm{rme}$, and the reflections in $\mathrm{c}Q$.
However, we reinitialize them due to them storing intermediate results which means we can reuse them.
In total, we need
\begin{multline}
    n_\mathrm{min}
    =
    \mathrm{min}(N, x\lceil\log_2 N\rceil) 
    +4\left\lceil
        \log_2 \max(N,M)
    \right\rceil
    +3
    \\
    +
    \left\lceil 
        \log_2 \varepsilon^{-1}
    \right\rceil
    +\max\left(
        \tilde{n}_E, 
        \tilde{n}^{(2a_V+1)}_\mathrm{Toffoli}, 
        \tilde{n}^{(\lceil\log_2 N\rceil+1)}_\mathrm{Toffoli}
    \right)
\end{multline}
qubits, which yields \cref{eq:memory}.
Here,  $\tilde{n}_E$ denotes the reusable ancilla qubits of the block encoding, $\tilde{n}^{(2a_V+1)}_\mathrm{Toffoli}$ the reusable ancilla qubits used in the reflections, and $\tilde{n}^{(\lceil\log_2 N\rceil+1)}_\mathrm{Toffoli}$ the reusable ancilla qubits used in the Toffoli gates of the Dicke state generation.
    
\subsubsection{Comparison to a classical approach}

\begin{figure}[h]
    \centering
    \includegraphics[width=0.95\linewidth]{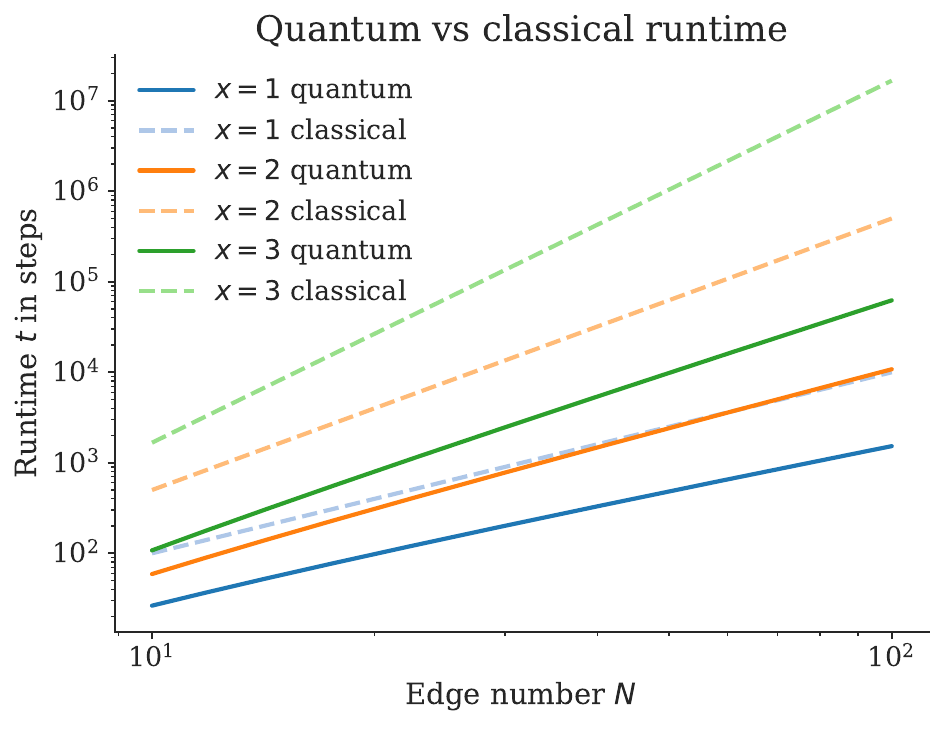}
    \caption{Double logarithmic plot showing runtime comparison between our \gls{ouralgo}, with runtime $t_\mathrm{min}=\mathcal{O}(\sqrt{N^x/x!}N\log\log N )$, and a state-of-the-art classical alternative based on the computation of the Frobenius distance for all $S$ Dicke states, with a runtime $t_\mathrm{cla}=\mathcal{O}(N^{x+1}/x!)$.
    Here, $N$ denotes the total number of edges in our graph and $x$ the number of inactive ones.
    For simplicity, we assume a constant precision $\log\varepsilon^{-1}$ for both the classical and quantum approach.}
    \label{fig:runtime}
\end{figure}

We compare the scaling behavior of $t_\mathrm{min}$ for constant $\varepsilon$ with that of a classical algorithm in \cref{fig:runtime}.
For the classical alternative we considered the brute force computation of all $S$ Frobenius distances $\{ \Dist(\tilde{d}) \}_{\tilde{d}}$, with a total runtime of $t_{\mathrm{cla}}=\mathcal{O}(N^{x+1}/x!)$.
We can see that our \gls{ouralgo} is clearly outperforming the classical approach for any $x \geq 1$.

Although in practice one might be able to utilize relaxation or other types of heuristics to improve upon the worst-case complexity of brute force approach, we choose this method to serve as a clear reference point to compare our approach to.
Note, that our approach is guaranteed to find the exact solution, which rules out comparison to classical approximate solvers.

In a classical approach, we could sort the configurations $\vec{\tilde{d}}$ in $\mathcal{O}(S)$ according to the corresponding squared Frobenius norm or rather distance 
\begin{align}
||\matr B - \matr B^\vec{\tilde{d}}||^2_\mathrm{F}= || \tilde{\matr B}||^2_\mathrm{F}=\sum_{r=0}^{M-1} \sum_{s=0}^{M-1} |\tilde{B}_{rs} |^2
\end{align}
where $\tilde{B}_{rs}$ are the elements of the matrix $\tilde{\matr B}$. Using \cref{eq:laplacian_definition}, we can rewrite the expression as 
\begin{align}
    ||\tilde{\matr B}||^2_\mathrm{F} =\sum_{r=0}^{M-1} \left(\sum_{l\in \mathcal{N}_r:\tilde{d}_{rl}=0} b_{rl} \right)^2 + 2 \sum_{(r,s)\in \overline{E}(\vec{\tilde{d}}) } |b_{rs}|^2
\end{align}
where $\overline{E}(\vec{\tilde{d}})$ is the set of inactive edges implying that $|\overline{E}(\vec{\tilde{d}})|=x$ given configuration $\vec{\tilde{d}}$. 
Hence, the costs and memory requirements are $\mathcal{O}(M)$ or $\mathcal{O}(N)$ and therefore $t_\mathrm{cla}=\mathcal{O}(SN)=\mathcal{O}(N^{x+1}/x!)$ for all configurations as we are working with sparse graphs and therefore sparse Laplacian matrices and $x \leq N$.
Hence, our \gls{ouralgo} requires similar or less memory than classical alternatives (cf. \cref{eq:memory}).

\subsection{Numerical Proof-of-Concept}
\label{sec:numerics}
\begin{figure}[h]
    \begin{center}
        \includegraphics[width=0.95\linewidth]{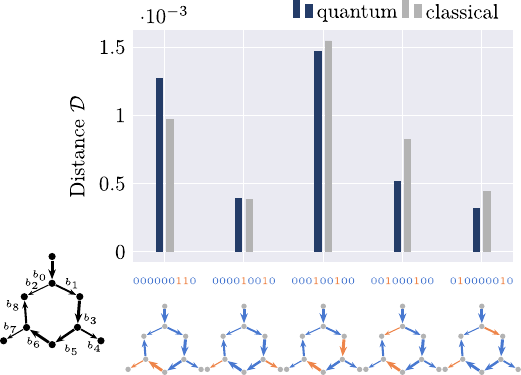}
    \end{center}
    \caption{%
        Distance calculated with our (quantum) approach compared to the classical approach for selected subgraph configurations of the IEEE-9 instance with two inactive edges ($x=2$) using $S_\mathrm{total}=1\mathrm{E}8$ shots in total. The reference graph is shown in black on the left side. For the subgraphs, green indicates the inactivity of the corresponding edge and blue the activity. The corresponding binaries $d_0 d_1 \dots d_8$ indicating the configuration i.e. topology of the subgraph are shown below.
    }\label{fig:numerics-histogram}
\end{figure}

As a numerical proof-of-concept of our algorithm, 
we focus on one of its central features: 
The creation of a superposition of quantum states whose amplitudes are related to the Frobenius distances $\mathcal{D}(\vec{\tilde{d}})$ (see \cref{eq:distance-definition}) for various configurations $\vec{\tilde{d}}$. 
To this end, 
we simulate the sampling from $\ket{\psi_\mathrm{f}}$ (see \cref{eq:psi2}) to calculate $\mathcal{D}(\vec{\tilde{d}})$ from the resulting histograms according to \cref{eq:frob_measurements}. 
As graph instances, we choose two weighted reference graphs - IEEE 4-bus \cite{grainger_4bus} and 9-bus \cite{anderson1997_9bus} systems representing electric power grids, given by a publicly available benchmarking suite \cite{pandapower.2018}. 
We compare our approach to the classical exact solution that calculates 
$\mathcal{D}(\vec{\tilde{d}})$ for all outage scenarios (i.e.~subgraphs) $\vec{\tilde{d}}$ out of $S$ for given $N$ and $x$.
In \cref{fig:numerics-histogram} we show the distances calculated for a selection of subgraphs of the IEEE-9 case and illustrate the different outage scenarios compared to the original graph.

\begin{figure}[h]
    \begin{center}
        \input{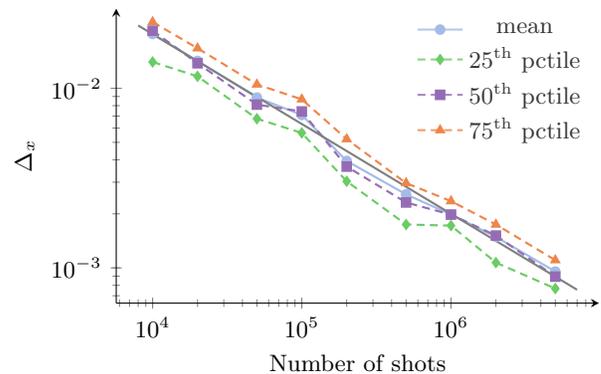}
    \end{center}
    \caption{%
        Precision of the quantum sampling approach given via the absolute difference $\Delta_x$ (see~\cref{eq:diff-quantum-classical-measure}) in the sum of calculated distances between our quantum method and the (exact) classical approach, against the number of samples from the quantum circuit simulation for the IEEE-4 instance with two inactive edges ($x=2$). Double logarithmic presentation of the axes are displayed. We show the mean of 10 different random seeds for the measurement simulation as well as different percentiles.
        The grey line indicates the expected behavior of $\mathcal{O}(1/\sqrt{S_\mathrm{total}})$ for $S_\mathrm{total}$ number of shots, as a guide to the eye.
    }\label{fig:numerics-diff-vs-numshots}
\end{figure}

By increasing the simulated number of samples, we can reproduce the exact results to arbitrary precision.
For this, we compare the absolute difference in the distance calculated by our quantum approach $\mathcal{D}$ and the classical (exact) approach $\mathcal{D}^0$ summed over all configurations with $x$ edges removed
\begin{equation}
    \Delta_x
    :=
    \sum_{\substack{%
        \vec{\tilde{d}}\\     
        |\vec{\tilde{d}}| = N-x}
    }
    \left|
    \mathcal{D}(\vec{\tilde{d}})
    -
    \mathcal{D}^0(\vec{\tilde{d}})
    \right|
    \label{eq:diff-quantum-classical-measure}
\end{equation}
Since $\mathcal{D}(\vec{\tilde{d}})$ is calculated from a finite number of shots $S_\mathrm{total}$, $\Delta_x$ is expected to vanish asymptotically as $\mathcal{O}(1/\sqrt{S_\mathrm{total}})$. 
This can be seen in~\cref{fig:numerics-diff-vs-numshots} which displays this quantity, for the IEEE-4 case for various number of samples, each averaged over 10 different random seeds in the simulation.
As expected, the precision increases with the number of samples.

Our implementation of this numerical proof-of-concept is publicly available under \href{https://jugit.fz-juelich.de/qugrids-public/topology-controlled-algorithm}{jugit.fz-juelich.de/qugrids-public/topology-controlled-algorithm}.
For the Dicke state preparation we employ \href{https://github.com/andre-juan/dicke_states_preparation}{code} based on \cite{bartschi2019deterministic} which prepares $\ket{D_x^N}$ in $\mathcal{O}(Nx)$. 
Additionally, we use the implementation of the \enquote{FABLE} algorithm~\cite{fable_code} based on \cite{camps2022_fable} for the block-encoding of the incidence matrix $\matr E$.

\section{Discussion}
\label{sec:discuss}
In this work we introduced \gls{ouralgo}, a quantum search algorithm for identifying the subgraph that is most similar to a weighted reference graph under a fixed number of edge removals. This task can be formulated as an NP hard \acrfull{ccbqp} which is computational demanding because the number of feasible configurations $S=\binom{N}{x}$ grows combinatorially. 
A key element of our approach is the encoding of all feasible configurations in a Dicke state, which is a superposition of computational basis states with fixed Hamming weight $x$. Another central feature is the association of the graph Laplacians with quantum states via matrix vectorization. The representation of the subgraph topologies in form of a Dicke state enables the application of controlled operations to flag the inactive edge states and, when combined with a block encoding of the reference graph's incidence matrix, the creation of a state whose amplitudes encode the Frobenius distances to the reference graph Laplacian. This construction can be used in two ways. First, in combination with \gls{ae} and Dürr’s minimum-finding algorithm, it provides a polynomial speedup compared to brute-force classical calculation for identifying the minimizing configuration. Second, the same state preparation procedure also provides the possibility to calculate quantities beyond the Frobenius distance. In particular, the resulting superposition carries states that can be used to evaluate quadratic forms of the type $\vec{a}^T \matr B^\vec{\tilde{d}} \vec{a}$ providing $\vec{a}$ as an additional input for the algorithm in form of a state $\ket{a}$. Possible applications of this feature have not been investigated in detail here and remain an interesting direction for future work. This observation further shows that the framework is not limited to a single optimization criterion. 
Taken together, these results suggest that \gls{ouralgo} provides a useful framework for constrained graph optimization tasks in which the subgraph topologies are determined by a fixed number of inactive edges $x$. It is important to distinguish our approach from heuristic quantum optimization strategies. \gls{ouralgo} is formulated as an exact algorithm and, on a fault-tolerant quantum computer, would return the exact solution minimizing the constrained problem. This distinguishes the present approach from alternative quantum optimization strategies based on QUBO formulations, where the cardinality constraint is incorporated, for example, through quadratic penalty terms. Such formulations generally do not guarantee to obtain the exact solution. A systematic comparison between the exact search-based approach developed here and heuristic methods are left for future work. To conclude: The present work may therefore serve as a starting point for exploring quantum approaches to a broader class of constrained graph optimization problems.

\section{Methods}
\label{sec:methods}
\subsection{Encoding grid topologies into quantum states}
\label{sec:topo-enc}
\noindent
The key idea of our approach lies in the evaluation
of a superposition of configurations $\Vec{d}$ which are encoded as quantum states taking the role of control qubits to perform further operations. In this section, we describe the creation of an uniform superposition of quantum states which can be interpreted as configuration states dictating the topology of the subgraph.

\subsubsection{Single edge removal}
\label{sec:single_edge_removal}
\noindent
In the case of a single edge removal, i.e.~$x=1$, the configuration is completely determined by indicating which of the $N$ edges is removed, i.e. by a number $j\in\{0, \dots, N-1\}$.
Hence, we can encode these configurations into a $n=\lceil \log_2 N \rceil$ qubit register via binary encoding
\begin{equation}
    j \to \ket{j}=\ket{j_0, \dots, j_{n-1}}
\end{equation}
with $j = \sum_{\ell=0}^{n-1} j_\ell 2^{\ell}$. 
Without loss of generality, we assume $N=2^n$ for the readability in the following.
An equal superposition of all $\ket{j}$ can be efficiently prepared via parallel Hadamard gates $H$:
\begin{equation}
    H^{\otimes n} \ket{0}^{\otimes n}
    =
    \frac{1}{\sqrt{2^n}} \sum_{j=0}^{2^n-1} \ket{j}
    \label{eq:topo_binary}
\end{equation}
We can interpret the decimal representation $\ket{j}$ in \cref{eq:topo_binary} as indicator for the non-operational status of edge $j$ which can be used as control register for controlled edge deactivation (see \cref{sec:topo}).

\subsubsection{Multiple edge removal}
\label{sec:dicke_states}
\noindent
For $x>1$ edges removed, one can extend the previous method by introducing additional registers -- one for each of the $x$ inactive edges labeled with $\set{j^{(k)}\mid \forall k \in [x]}$
\begin{equation}
    \frac{1}{\sqrt{2^n}} \sum_{j=0}^{2^n-1} \ket{j}
    \to
    \frac{1}{\sqrt{2^{xn}}} \sum_{j=0}^{2^n-1} \bigotimes_{k\in[x]}^x .\ket{j^{(k)}}
\end{equation}
However, in the following, we introduce another subgraph configuration encoding in $N$ qubits, that is fixed for any $x$ and scales slightly better for large $x=\mathcal{O}(N)$. For this we utilize Dicke states \cite{dicke1954coherence}, that are equal superpositions of computational basis states $\ket{d}\in \mathcal{D}_x^N=\{\ket{d}\in \mathbb{C}^{2^N} \mid \abs{\vec{d}}^2 = x\}$ with fixed Hamming weight $x$, defined as
\begin{subequations}
\begin{align}
    \ket{D_x^N} 
    &= 
    \binom{N}{x}^{-1/2} \mathrm{Perm}\left(  \ket{1}^{\otimes x} \otimes \ket{0}^{\otimes (N-x)} \right)\\
    &:=\frac{1}{\sqrt{S}}\sum_{d\in \mathcal{D}_x^N} \ket{d}\label{eq:dicke_short},
\end{align}
\end{subequations}
where $\mathrm{Perm}(\cdot)$ denotes the sum over all possible permutations.
The shorthand notation $\ket{d}$ in \cref{eq:dicke_short} can naturally be associated with the bitstring $\vec{\tilde{d}}=\vec{1}-\vec{d}$ which we refer to as \emph{configuration} or \emph{grid topology}. 
For example, with $N=4$ and $x=2$ we get
\begin{multline}
    \ket{D_2^4}
    =
    \frac{1}{\sqrt{6}} \left( 
        \ket{1100} + \ket{1010}+ \ket{1001}
    \right.
    \\
    \left.
        + \ket{0110}+ \ket{0101} + \ket{0011} 
    \right)
    \\
    =
    \frac{1}{\sqrt{6}}\left( \ket{12} + \ket{10} + \ket{9} + \ket{6} + \ket{5} + \ket{3} \right).
\end{multline}
Note that we use the little-endian convention in the last step. 

\subsection{Vectorizing matrices to define quantum states}
\label{sec:methods_vectorizing}
In this section, we provide the essential equations of graph theory that lay the foundation of our quantum algorithm. In particular, we demonstrate how the vectorized Laplacians can be associated with quantum states. 

The Laplacian $\matr B$ is related to the incidence matrix $\matr{E} \in \mathbb{R}^{M \times N}$ of graph $G$ with components
\begin{equation}
   E_{me} := \left\{
   \begin{array}{r l}
      + 1 & \; \mbox{if edge $e$ starts at node $m$},  \\
      - 1 & \; \mbox{if edge $e$ ends at node $m$},  \\
      0     & \; \mbox{otherwise}.
  \end{array} \right. 
  \label{eq:def_incidence}
\end{equation}
via 
\begin{align}
    \matr B&=\matr E \matr \Bd \matr E^T=\sum_{i=0}^{N -1} b_i \vec{v}_i \vec{v}_i^T,
\end{align}
where $\vec{v}_i$ is the $i$-th column vector of the incidence matrix $\matr E$ and $\matr \Bd = \mbox{diag}(b_0,\ldots,b_{N-1})$ the diagonal matrix whose entries correspond to the weights of the edges. 
In the following, we \emph{vectorize} the Laplacians,
resulting in a vector $\mathrm{vec}(\matr B)\in \mathbb{R}^{M^2}$. 
The vectorization $\mathrm{vec}(\matr A)$ of a matrix $\matr A \in \mathbb{R}^{p \times q}$ is obtained by stacking its column vectors $\vec{a}_i\in \mathbb{R}^p$ on top of each other so that
\begin{equation}
    \mathrm{vec}(\matr A)
    =
    \begin{pmatrix}
        \vec{a}_0 \\
        \vdots \\
        \vec{a}_{q-1}
    \end{pmatrix}
    =
    \sum_{i=0}^{q-1} \vec{e}_i \otimes \matr A \vec{e_i}
\end{equation}
with $\vec{e}_i$ denoting the $i$-th Euclidean basis vector. 
The vectorization of the diagonal matrix $\matr \Bd \in \mathbb{R}^{N \times N}$ 
\begin{equation}
    \matr \Bd=\sum_{i=0}^{N -1} b_i \vec{e}_i \vec{e}_i^T \xrightarrow[]{\mathrm{vec}}  \mathrm{vec}(\matr \Bd)=\sum_{i=0}^{N -1} b_i \vec{e}_i \otimes \vec{e}_i
\end{equation}
enables us to define the corresponding quantum states (up to normalization) as
\begin{align}
    \mathrm{vec}(\matr \Bd )=\sum_{i=0}^{N -1} b_i \vec{e}_i \vec{e}_i \longrightarrow \ket{\matr \Bd}&=\sum_{i=0}^{N -1} b_i \ket{i} \ket{i} ,\label{eq:state_diag}
\end{align}
where we omit the Kronecker product $\otimes$ as common practice in quantum mechanics. 
In our quantum algorithm, we prepare the quantum state $\ket{\matr \Bd}$ in \cref{eq:state_diag} via amplitude encoding in two quantum registers of size $n=\lceil \log_2 N\rceil$ each. 
If the total number of edges $N$ is no power of 2, we add $2^n-N$ 0's to the weight vector $\Vec{b}=(b_0, b_1,..., b_{N-1},0,...0)$ for the state preparation routine.

In the same manner, we associate the vectorized Laplacians
\begin{multline}
    \mathrm{vec}({\matr B})
    =
    \mathrm{vec} \left(\matr E \matr \Bd \matr E^T \right)
    =(\matr E \otimes \matr E) \mathrm{vec} (\matr \Bd)
    \\
    = \sum_{i=0}^{N -1}  b_i (\matr E \otimes \matr E) (\vec{e}_i  \otimes \vec{e}_i ) 
    =\sum_{i=0}^{N -1}  b_i \vec{v}_i \otimes \vec{v}_i
    \label{eq:vec_laplacian}
\end{multline}
to quantum states (up to normalization) of the form
\begin{align}
    \mathrm{vec}(\matr B)= \sum_{i=0}^{N -1}   b_i \vec{v}_i \vec{v}_i  \longrightarrow \ket{\matr B}&=\sum_{i=0}^{N -1} b_i \ket{v_i} \ket{v_i}.
    \label{eq:state_laplacian}
\end{align}
Note that we can write
\begin{align}
    \matr E \vec{e}_i
    =
    \vec{v}_i 
    =
    \Tilde{\vec{e}_r} 
    -  \Tilde{\vec{e}_s},
    \label{eq:incidence_mapping}
\end{align}
where $\vec{e}_i \in \mathbb{R}^{N}$ is the $i$-th basis vector of the edge space, $\Tilde{\vec{e}_r} \in \mathbb{R}^{M}$ the $r$-th basis vector of the node space.
The edges are denoted with $i=(r,s)$, where $r$ and $s$ are its input and output nodes, respectively.

As indicated in \cref{eq:vec_laplacian}, we aim to encode \cref{eq:state_laplacian} starting from \cref{eq:state_diag}.
However, the incidence matrix $\matr E$ is generally not unitary. 
Hence, the transformation shown in \cref{eq:vec_laplacian} cannot be performed on quantum hardware without modifications. 
We solve this problem by extending the registers with ancilla qubits and embed $\matr E $ in a larger unitary matrix -- a common procedure which is called block-encoding.

\subsection{Block encoding of the incidence matrix}
\label{sec:block_encoding}
\noindent
Quantum computing is limited to unitary operations.
However, non-unitary square matrices $\matr A$ can be embedded in larger unitary matrices $U_A$ via block-encoding (see e.g. \cite{wiebe2019} for an overview and its usage for quantum singular value transformation). We call $U_A$ an $(\alpha,a, \varepsilon)$ block encoding of a squared matrix $A \in \mathbb{R}^{N \times N}$ with $N=2^n$ if
\begin{align}
    \abs{
        \frac{\matr A}{\alpha}- \left(\bra{0}^{\otimes a} \otimes \mathds{1}_n \right) U_A \left(\ket{0}^{\otimes a} \otimes \mathds{1}_n \right)
    }
    \leq \varepsilon,
    \label{eq:std-block-encoding}
\end{align}
where $\alpha$ denotes the subnormalization factor, $a$ the number of ancilla qubits, and $\mathds{1}_n$ the identity operator of dimension $2^n$.
In our case, we should note that the incidence matrix $\matr E \in \mathbb{R}^{M \times N}$ is not necessarily square as  it maps the edge space $E$ to the vertex space $V$.
Further, neither $N$ nor $M$ are necessarily powers of 2.
Therefore, we introduce $\matr E_2$, which mapps the binary encodings of $\vec{e}_i$ and $\vec{v}_i$ onto each other
\begin{equation}
    \matr E_2 \ket{i}
    =
    \ket{v_i},
\end{equation}
where $\ket{i}\in \mathbb{C}^{2^n}$ and $\ket{v_i}\in \mathbb{C}^{2^m}$, with $n=\lceil\log_2 N\rceil$ and $m=\lceil\log_2 M\rceil$.
The corresponding block encoding $U_E$ features
\begin{align}
    \abs{
        \frac{\matr E_2}{\alpha}
        -
        \left(\bra{0}^{\otimes a_V} \otimes \mathds{1}_m \right) U_E \left(\ket{0}^{\otimes a_E} \otimes \mathds{1}_n \right),
    }
    \leq\varepsilon,
\end{align}
with $a_E+n=a_V+m$.
The normalization factor $\alpha=1/2$ necessary for sparse incidence matrices is constant if we have oracle access to $\matr E$~\cite{childsRelationshipContinuousandDiscreteTime2010, berryBLACKBOXHAMILTONIANSIMULATION2012}. The unitary $U_E$ is also a block encoding of the original $\matr E$.
In practice, we implement $U_E$ by extending $\matr E$ to $\matr E_\mathrm{s}\in \mathbb{R}^{K\times K}$ with $K = 2^{\max(n,m)}$ with zeros and use standard block encoding (see \cref{eq:std-block-encoding}) for $\matr E_\mathrm{s}$.

$\matr E$ is a sparse and highly structured matrix as its column vectors $\vec{v}_i$ only feature $0$ entries except for two entries which are $\pm 1$ i.e. $|E_{ij}| \leq 1 $ (recall \cref{eq:def_incidence}). 
This fulfills Theorem 1 of Ref.~\cite{camps2022_fable}, which limits the number of necessary ancilla qubits and we have $a_E+n=a_V+m=2\max(n,m)+1$.

In our case, we want to apply the block encoding of $\matr E$ to two registers. Therefore, we modify \cref{eq:state_laplacian} up to normalization according to
\begin{multline}
    \ket{\matr B}
    =
    U_{E^2}
    \sum_{i=0}^{N-1} b_i \ket{0^{\otimes a_E}, i} \ket{0^{\otimes a_E}, i}
    \\
    = 
    \sum_{i=0}^{N -1} b_i \frac{ \ket{0^{\otimes a_V}, v_i}  \ket{0^{\otimes a_V}, v_i} }{\alpha^2} + \ket{\mathrm{garbage}},
    \label{eq:state_after_BE}
\end{multline}
after extending the original registers with $a_E$ ancilla qubits each for the block encoding
\begin{align}
    U_{E^2} = U_E\otimes  U_E.
\end{align}
Note that $\ket{0^{\otimes a_E}, i}$ and $\ket{0^{\otimes a_V}, v_i}$ are of the same size.
Here, $\ket{\mathrm{garbage}}$ gathers all states, that are orthogonal to $\ket{0^{\otimes 2a_V}}$ i.e. it has at least one ancilla qubit in state $\ket{1}$.
The state $\ket{\matr B}$ can be understood as the quantum state associated with the vectorized Laplacian of the reference graph as it includes all edges in operational mode.  

\subsection{Topology controlled operations}
\label{sec:topo}
\noindent
Depending on the operational status of the edges as represented as $\ket{d}$, our goal is to set the corresponding weights to zero which implies skipping these terms in the summation of $\ket{\matr \Bd}$ or $\ket{\matr B}$. 
Therefore, we define controlled operations which project out or leave invariant the states associated with weights, depending on the operational status of the corresponding edges. 
Projectors are hermitian operators, but not necessarily unitary. Hence, we introduce flag qubits $\ket{\cdot}_\mathrm{f}$ labeled with index $\mathrm{f}$, to define unitary transformations.

\subsubsection{Single edge removal}
\noindent
We start with the $x=1$ case, in which the state $\ket{e}$ implies the non-operational status of edge $e$.
The unitary operation for the labeling and with it deactivation of the non-operational edges is denoted as $U_\mathrm{rse}$ (rse stands for "remove single edge"), which acts as follows:
\begin{multline}
   U_\mathrm{rse} \frac{1}{\sqrt{N}} \sum_{e,i=0}^{N-1} \ket{e}  \ket{i} \ket{0}_\mathrm{f} 
   \\
   =
   \frac{1}{\sqrt{N}} \sum_{e=0}^{N-1} \ket{e} \left(
        \sum_{i \neq e}^{N -1}\ket{i}  \ket{0}_\mathrm{f} 
        +  \ket{e}  \ket{1}_\mathrm{f} 
    \right) .
\end{multline}
Here, $\ket{1}_\mathrm{f}$ is a flag qubit needed to mark the state associated with the outaged edge.
The exact form of $U_\mathrm{rse}$ is
\begin{multline}
    U_\mathrm{rse} 
    =
    \sum_{i=0}^{N -1} 
    \ketbra{i}{i} \otimes \left(\mathds{1}_N- \ketbra{i}{i} \right) \otimes \mathds{1}_1 
    \\
    + 
    \sum_{i=0}^{N -1} \ketbra{i}{i} \otimes \ketbra{i}{i}  \otimes \sigma_x,
\end{multline}
where $\sigma_x$ is the unitary and hermitian Pauli x-matrix.
As a result, we are able to mark the states associated with the non-operational status of the edge with the flag qubit in state $\ket{1}_\mathrm{f}$.

\subsubsection{Multiple edge removal}
\label{sec:multi_controlled}
\begin{figure}[h!]
\centering
\input{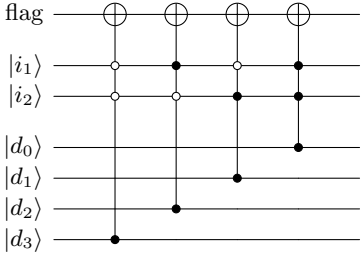}
\caption{
    Illustration of the topology controlled operation for $N=4$ via the multi controlled X-gate. Circuit corresponds to example presented in \cref{eq:ctrl_state_a,eq:ctrl_state_b,eq:ctrl_state_c,eq:ctrl_state_d} where $\ket{i}=\ket{i_1} \ket{i_2}$. 
}
\label{fig:multi_x_gate}
\end{figure}

\noindent
As stated in \cref{sec:dicke_states}, the Dicke state $\ket{D^N_x}$ is used to represent all $S$ possible grid configurations for $x>1$. 
To define the corresponding controlled unitary edge deactivation, we use the shorthand notation introduced in \cref{eq:dicke_short} where we label each state as  $\ket{d}=\ket{d_0 d_1 ... d_{N-1}}$. 
Its binaries $d_i$ indicate the operational status of edge $i$. If $d_i=1$, edge $i$ is removed. Recall that we use in \cref{eq:frob_class} $\tilde{d}_i=0$ of $\vec{\tilde{d}}$ indicating the inactivity of edge $i$.
The corresponding unitary transformation $U_\mathrm{rme}$ (rme stands for "remove multiple edges") has to satisfy
\begin{multline}
    U_\mathrm{rme} \ket{D^N_x} \sum_{i = 0}^{N-1} \ket{i} \ket{0}_\mathrm{f}
    \\
    =
    \frac{1}{\sqrt{S}} \sum_{d\in\mathcal{D}_x^N} \ket{d} \sum_{i = 0}^{N-1}\ket{i}
    \left( 
        (1-d_i) \ket{0}_\mathrm{f} 
        + d_i \ket{1}_\mathrm{f} 
    \right)
    \label{eq:controll_unit}.
\end{multline}
In the same manner as for the single edge removal case, we are able to mark all states, which are associated with the removed edges dictated by $\ket{d}$, with the flag qubit $\ket{1}_\mathrm{f}$. The following transformation has the desired action and is unitary 
\begin{multline}
    U_\mathrm{rme}  
    =
    \sum_{d\in\mathcal{D}_x^N} \ketbra{d}{d} \otimes \sum_{i=0}^{N-1}  \ketbra{i}{i} \otimes (1-d_i) \mathds{1}_1
    \\
    +
    \sum_{d\in\mathcal{D}_x^N} \ketbra{d}{d} \otimes \sum_{i=0}^{N-1} \ketbra{i}{i} \otimes d_i \sigma_x. 
    \label{eq:u_ctrl_dicke}
\end{multline}
This transformation can be realized via a combination of multi-controlled X-gates, which flip the flag qubit given control states determined by $\ket{d_i}\ket{i} $.
Let us illustrate the logic with a small example for $N=4$ (see \cref{fig:multi_x_gate}):
\begin{subequations}
    \begin{align}
    i=0 \to\mathrm{ flag\;if\;} \ket{d_0,i}=\ket{100}  
    \label{eq:ctrl_state_a}
    \\
    i=1\to\mathrm{ flag\;if\;} \ket{d_1,i}=\ket{110}    \label{eq:ctrl_state_b}
    \\
    i=2\to\mathrm{flag\;if\;} \ket{d_2, i}=\ket{101}    \label{eq:ctrl_state_c}
    \\
    i=3\to \mathrm{flag\;if\;} \ket{d_3, i} =\ket{111}
    \label{eq:ctrl_state_d}
    \end{align}
\end{subequations}
With this, we are able to perform controlled operations in a quantum parallel manner controlled by a superposition of topology configurations.

\subsection{State preparation}

\label{sec:state_prep}
\noindent
We describe in this section how to combine all the routines, introduced in the previous sections, to generate a quantum state which is used in a quantum routine to return a state $\ket{d}$, which can be associated to the subgraph configuration $\vec{\tilde{d}}$, with probability proportional to the Frobenius distance $\Dist(\vec{\tilde{d}})$ (see \cref{eq:frob_class}).
The first step is the initialization of the necessary registers to
\begin{align}
    \ket{\psi_\mathrm{i}}=\ket{0^{\otimes N}}_\mathrm{DS} \ket{0^{\otimes a_E}, 0^{\otimes n}}\ket{0^{\otimes a_E}, 0^{\otimes n}} \ket{0}_\mathrm{f},
\end{align}
which we categorise into three groups.

The first $N$-qubit register is for the subgraph configurations encoded in form of Dicke states.
The encoding of the latter is described in \cref{sec:dicke_states}.
Next to this, we require $a_E+n$ qubits twice for the encoding of the edge states $\ket{i}$ and the block encoding of the incidence matrix $\matr E$, which maps from the edge space to the node space (see \cref{sec:block_encoding}).
At last, the state $\ket{0}_\mathrm{f}$ is the flag qubit needed to mark the activation of the edges dictated by the qubits $\ket{d}=\ket{d_0 d_1 \dots d_{N-1}}$ of $\sum_{d\in \mathcal{D}_x^N} \ket{d}$. Flag qubit in $\ket{1}_\mathrm{f}$ implies removal of the corresponding edge indicated by $d_i=1$.
For a detailed analysis of the underlying controlled transformation $U_\mathrm{rme}$ see \cref{eq:controll_unit}.

The first step in the state preparation is the generation of the Dicke stats with $U_\mathrm{DS}$ applied to $\ket{0^{\otimes N}}_\mathrm{DS}$ so that
\begin{align}
    U_\mathrm{DS}\ket{0^{\otimes N}}_\mathrm{DS}
    =
    \ket{D^{N}_x}_\mathrm{DS} =\frac{1}{\sqrt{S}}\sum_{d \in \mathcal{D}_x^N} \ket{d}_\mathrm{DS}.
\end{align}
In parallel, we encode the edge weights $b_i$ via amplitude encoding $U_\mathrm{enc}$ (see below \cref{sec:methods}) according to
\begin{multline}
  U_\mathrm{enc} \ket{0^{\otimes a_E}, 0^{\otimes n} } \ket{0^{\otimes a_E}, 0^{\otimes n} } 
  \\
  =
  \frac{1}{\sqrt{W}}
  \sum_{i=0}^{N -1} b_i \ket{0^{\otimes a_E},i} \ket{0^{\otimes a_E},i} 
    =:\ket{\mathbf{b}} 
    \label{eq:weight-encoding}.
\end{multline}
Here, we introduce the weight normalization 
\begin{equation}
    W
    =
    \sum_{i=0}^{N-1} b_i^2 ,
\end{equation}
that is necessary for amplitude encoding. Then, we employ the topology controlled operations $U_\mathrm{rme}$, thus we have
\begin{multline}
   \ket{\psi_1}= \frac{1}{\sqrt{SW}}\sum_{d \in \mathcal{D}_x^N} \ket{d}_\mathrm{DS} \sum_{i=0}^{N-1} b_i \ket{0^{\otimes a_E}, i} \ket{0^{\otimes a_E}, i} 
   \\
   \otimes \left(
        (1-d_i) \ket{0}_\mathrm{f} 
        + d_i \ket{1}_\mathrm{f} 
    \right)  
\end{multline}
where $d_i$ is the $i$-th bit of $\ket{d}=\ket{d_0d_1...d_{N-1}}$. 
Next, the quantum circuit $U_{E^2}$ associated with the block-encoding of $\matr E$ transforms the previous into the final state
\begin{multline}
   \ket{\psi_\mathrm{f}} 
   =
   \frac{1}{\alpha^2\sqrt{SW}}
   \sum_{d \in \mathcal{D}_x^N} \ket{d}_\mathrm{DS} \sum_{i=0}^{N-1} b_i \ket{0^{\otimes a_V},v_i} \ket{0^{\otimes a_V}, v_i}
   \\ 
   \otimes\left((1-d_i) \ket{0}_\mathrm{f} + d_i \ket{1}_\mathrm{f} \right) 
   + \ket{\mathrm{garbage}} ,
   \label{eq:psi2}
\end{multline}
where the number of ancilla qubits in $\ket{0^{\otimes a_V}, v_i}$ went from $a_E$ to $a_V$. Following the notation of \cref{eq:frob_class} using $d_i=1-\tilde{d}_i$, we can rewrite $ \ket{\psi_\mathrm{f}}$ as
\begin{multline}
   \ket{\psi_\mathrm{f}} 
   =
   \frac{1}{\alpha^2\sqrt{SW}}
   \sum_{d \in \mathcal{D}_x^N} \ket{d}_\mathrm{DS} \sum_{i=0}^{N-1} b_i \ket{0^{\otimes a_V},v_i} \ket{0^{\otimes a_V}, v_i}
   \\ 
   \otimes\left(\tilde{d}_i \ket{0}_\mathrm{f} + (1-\tilde{d}_i) \ket{1}_\mathrm{f} \right) 
   + \ket{\mathrm{garbage}}.
\end{multline}
Note that the joint probability to measure a particular configuration $\ket{d}$ together with the flag qubit in state $\ket{1}_\mathrm{f}$ and the ancilla qubits in $\ket{0^{\otimes 2a_V}}$ matches the Frobenius distance $\Dist(\vec{\tilde{d}})$  (recall \cref{eq:frob_class})
\begin{align}
  \Dist(\vec{\tilde{d}})
  =
  \sum_{i,j=0}^{N-1} b_i b_j (1-\tilde{d}_i)(1-\tilde{d}_j)(\vec{v}_i^T \vec{v}_j)^2
\end{align}
up to constant factors which are fully determined by the normalization in the block encoding of $\matr E$, the encoding of $\ket{\mathbf{b}}$, and the Dicke state $\ket{D^N_x}$.
To make it clear: The probability $p(\ket{d} \ket{0^{\otimes 2a_V}}\ket{1}_\mathrm{f} )$ reads
\begin{multline}
   p(\ket{d} \ket{0^{\otimes 2a_V}}\ket{1}_\mathrm{f} )
   \\ 
    =
    \frac{1}{\alpha^4 SW} \sum_{i,j=0}^{N-1} b_i b_j  d_i d_j \abs{\braket{v_i | v_j}}^2
    \\
    =\frac{1}{\alpha^4 SW} \sum_{i,j=0}^{N-1} b_i b_j  (1-\tilde{d}_i) (1-\tilde{d}_j) \abs{\braket{v_i | v_j}}^2
    \\
    =
    \frac{ \Dist(\vec{\tilde{d}})}{\alpha^4  SW}
   \approx \frac{S_{\mathrm{succ}}^d}{S_\mathrm{total}}.
   \label{eq:frob_measurements}
\end{multline}
Here $S_\mathrm{succ}^d$ are the number of measurements yielding $\ket{d}$, $\ket{0^{\otimes 2a_V}}$ and $\ket{1}_\mathrm{f}$ whereby $S_\mathrm{total}$ is the total number of shots.

\subsection{Frobenius distance label generation with amplitude estimation}
\label{sec:amp-amp}

\noindent
We can determine the configuration $\vec{\tilde{d}}$, represented by the computational basis state $\ket{d} \in \mathcal{D}^N_x$ with Hamming weight $x$, with probability equal to the Frobenius distance $\Dist(\vec{\tilde{d}})$ up to constant factors by preparing $\ket{\psi_\mathrm{f}}$ and measuring $\ket{1}_\mathrm{f} \ket{0^{\otimes 2a_V}}$ in the ancilla registers. We propose an alternative method to avoid this inefficient sampling approach: First, we add another register $\ket{\cdot}_\mathrm{p}$ with $\lceil\log_2 \varepsilon^{-1}\rceil$ qubits.
This allows us to amplify the probability of measuring the state $\ket{d}$ with minimum Frobenius distance in a next step.

The tool we utilize for generating the label is \acrfull{ae}~\cite{brassard2000quantum}.
As the name suggests, it can estimate the amplitude of a chosen state -- in our case the state propotional to $\ket{\mathrm{s}} = \ket{1}_\mathrm{f}\ket{0^{\otimes 2a_V}}$ -- and returns it in a secondary quantum register.
Within this notation, we simplify $\ket{\psi_\mathrm{f}}$ and have
\begin{multline}
    \ket{\psi_\mathrm{f}}
    =
    \sum_{d \in \mathcal{D}_x^N} \ket{d}_\mathrm{DS} \left(
        \sqrt{p_d} \ket{\mathrm{V}}\ket{\mathrm{s}}
        +\sqrt{1-p_d}\ket{\mathrm{f}}
    \right),
    \\
    \ket{\mathrm{V}}
    =
    \frac{1
    }{
    \alpha^2\sqrt{SWp_d}
    }\sum_{i=0}^{N-1}b_i d_i\ket{v_i}\ket{v_i},
\end{multline}
where, $p_d = p(\ket{d}\ket{1}_\mathrm{f}\ket{0^{\otimes 2a_V}} )$.
All states, that are orthogonal to $\ket{\mathrm{s}}$ and less important in the following, are gathered in $\ket{\mathrm{f}}$. We denote the unitary operation generating $\ket{\psi_\mathrm{f}}$ from $\ket{D_x^N}\ket{0^{\otimes (2a_V+1)}}$ as described in \cref{sec:state_prep}
\begin{subequations}
\begin{align}
    \mathrm{c}U_\mathrm{f} 
    &=
    \sum_{d \in \mathcal{D}_x^N} \ketbra{d}{d}\otimes U_d,\\
    U_d\ket{0^{\otimes (2a_V+1)}}
    &=
    \sqrt{p_d} \ket{\mathrm{V}}\ket{\mathrm{s}}
    +\sqrt{1-p_d}\ket{\mathrm{f}}.
\end{align}
\end{subequations}

\gls{ae} requires two phase gates.
The first reflects around the success state
\begin{equation}
    P_{\mathrm{s}} =\mathds{1}_{2a_V+1} - 2\ketbra{\mathrm{s}}{\mathrm{s}},
\end{equation}
and can be implemented with a multi-controlled $Z$ gate, with control wires to $\ket{0^{\otimes 2a_V}}$ encapsulated by two NOT gates each, and target to 
$\ket{1}_\mathrm{f}$.

The second phase gate reflects around the state $U_d\ket{0^{\otimes (2a_V+1)}}$
\begin{equation}
    P_d
    =
    U_dP_0 U_d^\dagger,
\end{equation}
with $P_0$ being the reflection around the initial state $\ket{0^{\otimes (2a_V+1)}}$, also implemented with a multi-controlled $Z$ gate, but this time encapsulated fully by NOT gates.
In what follows, we need the controlled version of it
\begin{equation}
    \mathrm{c}P_\mathrm{f}
    =
    \sum_{d \in \mathcal{D}_x^N} \ketbra{d}{d}\otimes P_d
    =
    \mathrm{c}U_\mathrm{f}(\mathds{1}_N\otimes P_0)\mathrm{c}U_\mathrm{f}^\dagger
\end{equation}

The combination of those two gates
\begin{equation}
    \mathrm{c}Q
    =
    - \mathrm{c}P_\mathrm{f} (\mathds{1}_N \otimes P_{\mathrm{s}})
    \\
    =
    \sum_{d \in \mathcal{D}_x^N} \ketbra{d}{d} \otimes (-P_dP_\mathrm{s})
\end{equation}
has the convenient eigenvalues
\begin{equation}
     \lambda_{\pm,d}  
     =
     e^{\pm 2 i \arcsin \sqrt{p_d}}.
\end{equation}

The next step in \gls{ae} is the encoding of the phase $\pm\vartheta_d =\pm 2\arcsin\sqrt{p_d}$ of $\mathrm{c}Q$ via \gls{qpe} in the secondary register $\ket{0^{\otimes a_\varepsilon}}_\mathrm{p}$ with $a_\varepsilon = \lceil\log_2 \varepsilon^{-1}\rceil$. To be more precise, we apply the \gls{qpe} for $\mathrm{c}Q$ to the Dicke state and have
\begin{multline}
    \mathrm{\gls{qpe}}\sum_{d \in \mathcal{D}_x^N} \ket{d}_\mathrm{DS} \ket{0^{\otimes (2a_V+1)}}\ket{0^{\otimes a_\varepsilon}}_\mathrm{p}
    \\
    =
    \sum_{d \in \mathcal{D}_x^N}\ket{d}_\mathrm{DS} 
    \left(
        \ket{\Psi_{+,d}} \ket{\vartheta_d}_\mathrm{p}
        +\ket{\Psi_{-,d}} \ket{-\vartheta_d}_\mathrm{p}
    \right),
    \label{eq:qpe}
\end{multline}
where $\ket{\Psi_{\pm,d}}$ are the eigenstates of $P_dP_\mathrm{s}$ and normalized for \cref{eq:qpe}.

For readability, we assumed without loss of generality, that $\vartheta_d$ can be encoded exactly in $a_\varepsilon$ qubits without specifying the encoding. The last simplification we do is mapping $\ket{\pm\vartheta_d}$ onto $\ket{\mathrm{sgn_b}(\pm\vartheta_d),\vartheta_d}$ -- from the so-called two's complement representation to the the sign-magnitude form -- with $\mathrm{sgn_b}(x\geq 0) = 0$ and $\mathrm{sgn_b}(x< 0) = 1$.
For this, we decrease the binary value by one with a subtraction circuit, that is controlled by the most significant qubit.
This returns the so-called one's complement form.
At this stage, a series of CNOT gates brings it into the sign-magnitude form.
All those gates are controlled by the most significant qubit of the phase register $\ket{\cdot}_\mathrm{p}$ and applied to the remaining qubits.

Applying this to our state returns
\begin{multline}
    \ket{\psi_\mathrm{label}}
    \\
    =
    \sum_{d \in \mathcal{D}_x^N} 
        \ket{d}_\mathrm{DS} 
        \left(
            \ket{\Psi_+} \ket{0}_\mathrm{p_0}
            +\ket{\Psi_-} \ket{1}_\mathrm{p_0}
        \right)
        \ket{\vartheta_d}_\mathrm{label},
    \label{eq:labeling}
\end{multline}
where we split the phase register according to $\ket{0^{\otimes a_\varepsilon}}_\mathrm{p} \to \ket{0}_\mathrm{p_0}\ket{0^{\otimes a_\varepsilon-1}}_\mathrm{label}$.
Now, $\ket{\psi_\mathrm{label}}$ is the combination of all Dicke states labeled with the phase $\vartheta_d = 2\arcsin\sqrt{p_d}$, which contains the Frobenius distance. The steps described above can be used for arbitrary configuration $y$ as input by replacing the initial superposition $\ket{D_x^N}$ with $\ket{y}$.
This yields
\begin{equation}
    \ket{\psi_\mathrm{label}^y}
    =
    \ket{y}
    \left(
        \ket{\Psi_+} \ket{0}_\mathrm{p_0}
        +\ket{\Psi_-} \ket{1}_\mathrm{p_0}
    \right)
    \ket{\vartheta_y}_\mathrm{label}.
\end{equation}

\subsection{Identifying the state with minimal Frobenius distance}
Labeling the computational basis states of the Dicke state superposition together with their squared Frobenius distance encoded in a quantum state is not sufficient to find the one with minimum distance in a feasible runtime.
We would still need to sample from a superposition of $S$ states.
However, we can amplify the amplitudes of the states with small $\vartheta_d$ and with it also small Frobenius distance, while damping the amplitudes of states with large $\vartheta_d$.
For this, we rely on a minimum finder introduced by Dürr and H\o yer~\cite{durr1996quantum}, which we will review in this section.

The minimum finder starts by randomly choosing a configuration $\vec{\tilde{y}} \in \{0;1\}^N$ out of $S$-many represented by a computational basis state $\ket{y}$ with Hamming weight $x$. 

Next, five steps\footnote{The original work mentions only three steps. Here, we separated the first step into three.} are repeated until a total runtime of $22.5\sqrt{S}+1.4\log^2 S$ is surpassed.
\begin{enumerate}
    \item Initialize $\ket{\psi_\mathrm{label}}$.
    \item Initialize $\ket{\psi_\mathrm{label}^y}$.
    \item Mark states with $\vartheta_d < \vartheta_y$.
    \item Apply quantum search algorithm
    \item Measure the label register and replace $y$ with the outcome if it is smaller than $y$.
\end{enumerate}
The implementation of the first two steps are described in the previous sections.
The third step requires a quantum comparison between the values stored in the registers $\ket{\vartheta_d}_\mathrm{label}$ and $\ket{\vartheta_y}_\mathrm{label}$ followed by a base flip of an ancilla qubit if $\vartheta_d< \vartheta_y$.
This can be achieved via a binary subtraction followed by an addition (see Appendix in Ref.~\cite{danz2025quantumoraclesfiniteelement} for more details).
Here, it  is assumed, that the quantum search algorithm, in step four, runs for at most $\mathcal{O}(\sqrt{S})$ iterations~\cite{boyerTightBoundsQuantum1998}.
At the end of this loop, we measure $\ket{y}$ representing the configuration $\vec{\tilde{y}}$ with the lowest Frobenius distance $\Dist(\vec{\tilde{y}})$ with high probability.

\subsection{Calculation of quadratic forms}
\label{sec:quad_from}
\noindent
One of the key features of our approach is the association of the vectorized Laplacians to quantum states (see \cref{eq:state_diag,eq:state_laplacian}). This formulation enables us to calculate further quantities beyond $||\matr B-\matr B^\vec{\tilde{d}}||_\mathrm{F}^2$, if we provide another real vector $\vec{a} \in \mathbb{R}^{M}$ as input of the algorithm. In this case, we can calculate (energy functional like) quadratic forms
\begin{subequations}
    \begin{align}
    \vec{a}^T \matr B^\vec{\tilde{d}} \vec{a}
    &=\mathrm{Tr}(\matr B^\vec{\tilde{d}} \vec{a} \vec{a}^T)\\
    &=\mathrm{vec}(\matr B^\vec{\tilde{d}})^T \mathrm{vec}(\vec{a}\vec{a}^T)\\
    &=\mathrm{vec}(\matr B^\vec{\tilde{d}})^T \vec{a} \otimes \vec{a} \label{eq:quad_from_iden}\\
    &= \sum_{i=0}^{N-1} b_i \tilde{d}_i  (\vec{a}^T \vec{v_i})^2     \label{eq:quad_form} 
\end{align}
\end{subequations}
\noindent
where we use the fact that $\matr B$ (and $\matr B^\vec{\tilde{d}}$) is real and symmetric. As stated before, we can create the state $\ket{\psi_\mathrm{f}}$ (see \cref{eq:psi2}) whose amplitudes can be associated with $\mathcal{D}(\vec{\tilde{d}})$. But we can also make use of the other states in this superposition as $\ket{\psi_\mathrm{f}}$ carries
\begin{multline}
\ket{\Phi_0}_\vec{d} = \frac{1}{\alpha^2\sqrt{SW}} \sum_{i=0}^{N-1} b_i (1-d_i) \ket{0^{\otimes a_V},v_i}\ket{0^{\otimes a_V},v_i} \ket{0}_\mathrm{f} \\
=\frac{1}{\alpha^2\sqrt{SW}} \sum_{i=0}^{N-1} b_i \tilde{d}_i \ket{0^{\otimes a_V},v_i}\ket{0^{\otimes a_V},v_i} \ket{0}_\mathrm{f} .
\end{multline}
Relating $\vec{a} \otimes \vec{a}$ to $\ket{0^{\otimes a_V},a} \ket{0^{\otimes a_V},a} \ket{0}_\mathrm{f} $ (up to normalization) and $\mathrm{vec}(\matr B^\vec{\tilde{d}})$ to $\ket{\Phi_0}_\vec{d}$ of $\ket{\psi_\mathrm{f}}$ allows us to calculate the quadratic form between Laplacians and real vectors
\begin{multline}
    \bra{\Phi_0}_\vec{d} \left( \frac{\ket{0^{\otimes a_V},a} \otimes\ket{0^{\otimes a_V},a} \ket{0}_\mathrm{f}}{|| \ket{a}||^2} \right)\\
    = \frac{1}{\alpha^2 \sqrt{SW}}\sum_{i=0}^{N-1} b_i \tilde{d}_i \frac{ \braket{v_i|a}^2}{|| \ket{a}||^2  } \\
    =\frac{\vec{a}^T \matr B^\vec{\tilde{d}} \vec{a}}{\alpha^2 \sqrt{SW}   ||\vec{a} ||^2}
\end{multline}
which is the "quantum analog" to \cref{eq:quad_form}. This feature can be used to calculate Dirichlet energy functional like quantities. Such forms are common in all kind of fields and also play a crucial role in the context of signal processing and state estimation of power systems \cite{dabush2023state}.

\section*{Data availability}
Data sets are available at \href{https://jugit.fz-juelich.de/qugrids-public/topology-controlled-algorithm}{jugit.fz-juelich.de/qugrids-public/topology-controlled-algorithm}.

\section*{Code availability}
Code available at \href{https://jugit.fz-juelich.de/qugrids-public/topology-controlled-algorithm}{jugit.fz-juelich.de/qugrids-public/topology-controlled-algorithm}.

\section*{Acknowledgments}
\noindent
 RK was funded by the project ``Quantum-based Energy Grids (QuGrids)", which is receiving funding from the programme ``Profilbildung 2022", an initiative of the Ministry of Culture and Science of the State of North Rhine-Westphalia.
SD were funded by the German Federal Ministry of Research, Technology and Space (BMFTR) in the project QUantum Algorithms to SImulate MAny-body Physics (QuASi-MaP, Grand No. 13N17336).
TS was funded by the German Federal Ministry of Research, Technology and Space (BMFTR) in the project quantum artificial intelligence for the automotive value chain (QAIAC), Funding No.~13N17166.

\section*{Author contributions}
\noindent
The project and algorithm was conceived and worked out by RK. RK and TS developed the problem formulation. RK and SD were responsible for the design of the algorithm and the technical details. SD provided the complexity analysis and resource estimation. Code was developed by RK. RK and TS contributed to the numerical proof-of-concept. TS provided overall scientific guidance. AB contributed the application-oriented conceptualization of the project within the power-systems domain. RK, SD, and TS are responsible for the writing and presentation of the manuscript.

\section*{Competing Interests}
The authors declare no competing interests.

\bibliography{ref}

\bibliographystyle{ieeetr} 

\vfill

\end{document}